\documentclass[%
 reprint,
 amsmath,amssymb,
 aps,
 prl,
]{revtex4-1}

\usepackage{graphicx}
\usepackage{dcolumn}
\usepackage{bm}
\usepackage{makecell}  
\usepackage{amsmath}   
\allowdisplaybreaks[4] 

\usepackage[
margin=0.71in,
]{geometry}

\begin{document}

\preprint{APS/123-QED}

\title{Orbital-angular-momentum-based experimental test of Hardy's paradox for multisetting and multidimensional systems}
\author{Dongkai Zhang}
\author{Xiaodong Qiu}
\author{Tianlong Ma}
\author{Wuhong Zhang}
\author{Lixiang Chen}
\email{chenlx@xmu.edu.cn}
\affiliation{Department of Physics and Collaborative Innovation Center for Optoelectronic Semiconductors and Efficient Devices, Xiamen University, Xiamen 361005, China}

\date{\today}

\begin{abstract}
Characterizing high-dimensional entangled states is of crucial importance in quantum information science and technology. Recent theoretical progress has been made to extend the Hardy's paradox into a general scenario with multisetting multidimensional systems, which can surpass the bound limited by the original version. Hitherto, no experimental verification has been conducted to verify such a Hardy's paradox, as most of previous experimental efforts were restricted to two-dimensional systems. Here, based on two-photon high-dimensional orbital angular momentum (OAM) entanglement, we report the first experiment to demonstrate the Hardy's paradox for multiple settings and multiple outcomes. We demonstrate the paradox for two-setting higher-dimensional OAM subspaces up to $d$ = 7, which reveals that the nonlocal events increase with the dimension. Furthermore, we showcase the nonlocality with an experimentally recording probability of 36.77\% for five-setting three-dimensional OAM subspace via entanglement concentration, and thus showing a sharper contradiction between quantum mechanics and classical theory.
\end{abstract}

\pacs{Valid PACS appear here}
\maketitle


\section{I. INTRODUCTION}

In 1935 Einstein, Podolsky, and Rosen (EPR) raised a famous paradox concerning the completeness of quantum mechanics \cite{Einstein35}. In 1964 Bell formulated the Bell's inequality to resolve the EPR paradox, which stated that the results of quantum theory could not be reproduced with a classical, deterministic local model based on  ``elements of reality '' \cite{Bell64}. Since then, numerous experiments have been performed to demonstrate the violation of Bell inequalities \cite{Brunner14}. An unsatisfactory feature in the derivation of Bell inequalities is that it applies only to statistical measurement procedures. In 1990s Hardy presented another logic paradox challenging the idea of locality and hidden variables \cite{Hardy92,Hardy93}, which represented an attempt to demonstrate nonlocality without inequalities, and, as such, Mermin referred to it as ``the best version of Bell's theore'' \cite{Mermin95}. Hereafter, a variety of versions of Hardy's paradox have been reported to increase the probability for demonstrating ``nonlocality without inequalities''. Boschi and co-workers developed the ladder version of Hardy's paradox for two spin-1/2 particles, significantly increasing the probability of the nonlocal events \cite{Boschi97}. It was also theoretically proved by Rabelo \cite{Rabelo12} that there exists an analogue of Tsirelson's bound \cite{Tsirelson80} for Hardy's test of nonlocality, irrespective of the dimension of the system, which was subsequently confirmed in experiment \cite{Chen17}. Recent progress was also made to generalize the Hardy's paradox into a most general framework of $n$-qubit system \cite{Jiang18}. While in the experimental implementation, the photonic polarization, energy-time, orbital angular momentum (OAM) \cite{Giuseppe97,Vallone11,Chen12,Karimi14,Luo18,Yang19} have been employed to test the Hardy's paradox, but only two-dimensional state spaces were considered.

High-dimensional systems (qudits) can provide higher information density coding, improve security in quantum communication, simplify the implementation of quantum logic, and inspire novel quantum imaging techniques \cite{Wang05,Bechmann20,Lanyon09,Erhard18,Qiu19}. The nonlocal feature of high-dimensional OAM entangled state was verified by using the generalized Bell inequalities \cite{Vaziri02,Dada11}. Unlike using inequalities, Chen \emph{et al.} formulated another new logical structure for two-qudit entangled states and showed that the probability of nonlocal events could grow significantly with the dimension $d$ \cite{Chen13}. Very recently, they further generalized the high-dimensional Hardy's proof to a ladder version, i.e., Hardy's paradox for general $(k, d)$ systems, where $(k, d)$ denotes a measuring scenario with $k$ settings and $d$ outcomes (dimensions) \cite{Meng18}. Such a generalization to a high-dimensional system is of crucial importance, because it is much closer to the original EPR scenario where the measurements have an arbitrarily large number of outcomes \cite{Vértesi10}. Besides, it can make the contradiction between quantum mechanics and local variable theory sharper than previous versions \cite{Rabelo12,Kunkri05,Seshadreesan11}. However, these theoretical schemes for both high-dimensional systems \cite{Chen13} and general $(k, d)$ systems \cite{Meng18} have not yet been translated into experimental implementations. Here we exploit the photonic orbital angular momentum (OAM) \cite{Allen92} to demonstrate the general multisetting multidimensional version of Hardy's paradox. In theory, we transform the Hardy's logical proof to an experiment-friendly model. In experiment, we employ two-photon high-dimensional OAM states generated by spontaneous parameter down-conversion (SPDC) and demonstrate the paradox for two-setting higher-dimensional OAM subspaces up to $d$ = 7. Our experimental observations reveal that the nonlocal events can increase with the dimension of system, significantly surpassing the bound limited by the original version \cite{Rabelo12}. Furthermore, we showcase the nonlocality with an experimentally recording probability of 36.77\% for five-setting three-dimensional OAM subspace $(5, 3)$ by entanglement concentration, which generalizes the ladder version of Hardy's proof \cite{Boschi97} to a truly high-dimensional scenario, showing a sharper contradiction between quantum mechanics and classical theory.

\section{II. THEORETICAL SCHEME}

Let us first summarize the generalized Hardy's paradox presented in \cite{Chen13,Meng18}. Consider a general $(k, d)$ scenario with two observers, Alice and Bob, each of them chooses $k$ set of measurements, on which the measurement outcomes range from 1 to $d$. And their von Neumann measurements are defined as, $\left| {A_s^i} \right\rangle \left\langle {A_s^i} \right|$ and $\left| {B_t^j} \right\rangle \left\langle {B_t^j} \right|$, respectively, where $i,j \in \left\{ {1,{\rm{ }}2,{\rm{ }}...{\rm{ }},{\rm{ }}k} \right\}$ and $s,t \in \left\{ {1,{\rm{ }}2,{\rm{ }}...{\rm{ }},{\rm{ }}d} \right\}$. In the paradox, it assumes the following zero probabilities:
\begin{eqnarray}
&P\left( {{A^1} < {B^k}} \right) = 0,\\
&P\left( {{B^{i - 1}} < {A^{i - 1}}} \right) = 0,  \quad{\rm{ for }}\;i = 2,3,...,k,\\
&P\left( {{A^i} < {B^{i - 1}}} \right) = 0,\quad{\rm{ for }}\;i = 2,3,...,k,
\end{eqnarray}
where $P\left( {{A^i} < {B^j}} \right) = \sum_{s < t} {P\left( {A_s^i,B_t^j} \right)} $ denotes the total joint probabilities in all cases that the measurement outcome of $A^i$  is strictly smaller than that of $B^j$. Within any local hidden variable theory, following Eqs. (1), (2) and (3), we can straightforwardly obtain an exactly zero probability, namely, $P\left( {{A^k} < {B^k}} \right) = 0$. However, with a suitable choice of measurements, quantum mechanics allows a non-zero probability,
\begin{eqnarray}
P\left( {{A^k} < {B^k}} \right) > 0.
\end{eqnarray}

As the OAM eigenstates form an orthogonal and complete basis \cite{Molina01}, twisted photons are the ideal candidate to realize a high-dimensional Hilbert space. Here, based on two-photon OAM entanglement generated by SPDC, we can translate the above theoretical strategy into an experimental implementation. In SPDC, high-dimensional OAM entangled state can be written as, ${\left| \Psi  \right\rangle _{\rm{SPDC}}} = \sum\nolimits_\ell  {{C_\ell }{{\left| \ell  \right\rangle }_A}{{\left| { - \ell } \right\rangle }_B}} $, where ${C_\ell }$ indicates the probability amplitude of finding one signal photon (A) with $\ell \hbar $ OAM and its partner idler photon (B) with $ - \ell \hbar $ OAM \cite{Mair01}. As we aim to explore a larger but finite subspace to formulate the Hardy's proof with OAM, we assume the optimal Hardy states, ${\left| \Psi  \right\rangle _{\rm{Hardy}}} = \sum\nolimits_{i = 1}^d {{c_{{\ell _i}}}{{\left| {{\ell _i}} \right\rangle }_A}{{\left| { - {\ell _i}} \right\rangle }_B}} $, in a specific $d$-dimensional OAM subspace. Thus we need to apply suitable entanglement concentration first to prepare ${\left| \Psi  \right\rangle _{\rm{Hardy}}}$ from ${\left| \Psi  \right\rangle _{\rm{SPDC}}}$, and look for all desired measurement bases, $\left| {A_s^i} \right\rangle \left\langle {A_s^i} \right|$ and $\left| {B_t^j} \right\rangle \left\langle {B_t^j} \right|$, to compute the optimal Hardy fraction. For mathematical convenience, we can rewrite the state as a diagonal matrix, $H = \rm{diag}{\rm{ }}({c_{{\ell _1}}},{c_{{\ell _2}}}, \ldots ,{c_{{\ell _d}}}).$ Our experimental situation is unlike the original proposal in \cite{Chen13,Meng18}, in which they started from two presetting standard bases for Alice and Bob, then calculated the desired two-photon entangled states, which were represented by upper (or lower) triangular matrices and could not be produced by SPDC. In the OAM space, we define the bases, $\left| {A_s^i} \right\rangle  = \sum\nolimits_{m = 1}^d {a_{s,m}^i\left| {{\ell _m}} \right\rangle } $ and $\left| {B_t^j} \right\rangle  = \sum\nolimits_{n = 1}^d {b_{t,n}^j\left| {{\ell _n}} \right\rangle } $ to specify the von Neumann measurements, where $\left\langle {A_s^i} \right.\left| {A_{s'}^i} \right\rangle  = {\delta _{ss'}}$ and $\left\langle {B_t^j} \right.\left| {B_{t'}^j} \right\rangle  = {\delta _{tt'}}$ within the $i$-th and $j$-th sets of measurements. In order to achieve the maximal successful probability, we need calculate the optimal weight amplitudes, $a_s^i = {\left[ {a_{s,1}^i,a_{s,2}^i, \ldots ,a_{s,d}^i} \right]^T}$ and $b_t^j = {\left[ {b_{t,1}^j,b_{t,2}^j, \ldots ,b_{t,d}^j} \right]^T}$, according to Eqs. (1), (2) and (3). We first consider Eq. (1), i.e., $P\left( {{A^1} < {B^k}} \right) = 0$ , which means that for all $s < t$, each $P\left( {A_s^1,B_t^k} \right)$ should be exactly zero, i.e., $\left| {B_t^k} \right\rangle  \bot {H^T}\left| {A_s^1} \right\rangle $, where $ \bot $ denotes the orthogonality symbol. Meanwhile, because of the mutual orthogonality,  $\left| {B_t^k} \right\rangle  \bot \left| {B_{t'}^k} \right\rangle \quad (t \ne t')$, we can obtain the following orthogonality relations,
\begin{widetext}
\begin{subequations}
\begin{align}
&\left| {B_d^k} \right\rangle  \bot \left\{ {{H^T}\left| {A_1^1} \right\rangle ,{H^T}\left| {A_2^1} \right\rangle ,{H^T}\left| {A_3^1} \right\rangle , \cdot  \cdot  \cdot ,{H^T}\left| {A_{d - 2}^1} \right\rangle ,{H^T}\left| {A_{d - 1}^1} \right\rangle } \right\}{\rm{,}}\\
&\left| {B_{d - 1}^k} \right\rangle \bot \left\{ {{H^T}\left| {A_1^1} \right\rangle ,{H^T}\left| {A_2^1} \right\rangle ,{H^T}\left| {A_3^1} \right\rangle , \cdot  \cdot  \cdot ,{H^T}\left| {A_{d - 2}^1} \right\rangle ,\left| {B_d^k} \right\rangle } \right\},\\
&\qquad\qquad\qquad\qquad\qquad\qquad\vdots \nonumber\\
&\left| {B_3^k} \right\rangle  \bot \left\{ {{H^T}\left| {A_1^1} \right\rangle ,{H^T}\left| {A_2^1} \right\rangle ,\left| {B_d^k} \right\rangle , \cdot  \cdot  \cdot ,\left| {B_5^k} \right\rangle ,\left| {B_4^k} \right\rangle } \right\},\\
&\left| {B_2^k} \right\rangle  \bot \left\{ {{H^T}\left| {A_1^1} \right\rangle ,\left| {B_d^k} \right\rangle ,\left| {B_{d - 1}^k} \right\rangle , \cdot  \cdot  \cdot ,\left| {B_4^k} \right\rangle ,\left| {B_3^k} \right\rangle } \right\},\\
&\left| {B_1^k} \right\rangle  \bot \left\{ {\left| {B_d^k} \right\rangle ,\left| {B_{d - 1}^k} \right\rangle ,\left| {B_{d - 2}^k} \right\rangle , \cdot  \cdot  \cdot ,\left| {B_3^k} \right\rangle ,\left| {B_2^k} \right\rangle } \right\}.
\end{align}
\end{subequations}
\end{widetext}
\begin{figure}[t]
\centerline{\includegraphics[width=1\columnwidth]{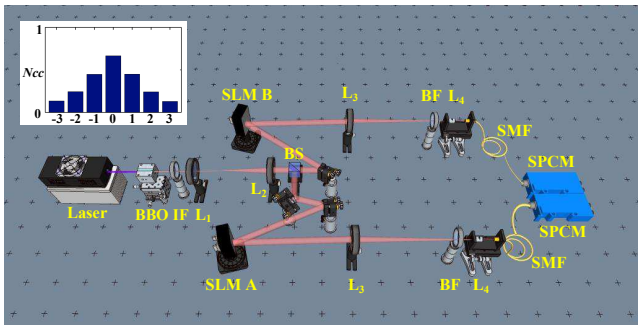}}
\caption{ (Color online) Experimental setup for demonstrating multisetting multidimensional Hardy's paradox with high-dimensional OAM entanglement. The inset is the experimentally measured two-photon OAM spectrum.}
\end{figure}

Since $H$ is of full rank and $\left\langle {A_s^1} \right.\left| {A_{s'}^1} \right\rangle  = {\delta _{ss'}}$, we know that the elements in the right-hand side of Eq. (5a) are linearly independent. Mathematically, $\left| {B_d^k} \right\rangle $ can then be uniquely determined by the entries of ${A^1}$. We can further determine $\left| {B_{d - 1}^k} \right\rangle$ based on Eq. (5b) together with the known $\left| {B_d^k} \right\rangle$. Along this line, we can determine all other $\left| {B_t^k} \right\rangle$ for $t = d - 2, \cdot  \cdot  \cdot ,1.$ Thus the $k$ set of measurements, ${B^k} = \{ \left| {B_1^k} \right\rangle ,\left| {B_2^k} \right\rangle , \cdot  \cdot  \cdot ,\left| {B_d^k} \right\rangle \} $, for photon B can all be calculated. We proceed to consider Eq. (2) with $i = 2$, namely, $P\left( {{B^1} < {A^1}} \right) = 0$, from which we can calculate $B^1$ from $A^1$ in a similar way. Then, by further consider Eq. (3) with $i = 2$, namely, $P\left( {{A^2} < {B^1}} \right) = 0$, we can also calculate $A^2$ from the known $B^1$. Along this line based on the ladder derivation of Eqs. (2) and (3), we can uniquely determine all other sets of measurements, ${A^3},{A^4}, \cdot  \cdot  \cdot ,{A^k}$ and ${B^2},{B^3}, \cdot  \cdot  \cdot ,{B^{k - 1}}.$ In other words, all the sets of the desired OAM measurement states, $A^i$ and $B^j$, can be determined finally. Therefore, we can calculate from Eq. (4) the successful probability,
\begin{eqnarray}
P\left( {{A^k} < {B^k}} \right) = \sum\limits_{s = 1}^{d - 1} {\sum\limits_{t = s + 1}^{d} {{{\left| {\left\langle {\Psi }
 \mathrel{\left | {\vphantom {\Psi  {A_s^k}}}
 \right. \kern-\nulldelimiterspace}
 {{A_s^k}} \right\rangle \left| {B_t^k} \right\rangle } \right|}^2}} }.
\end{eqnarray}

Remember that in Eq. (6), the sets of measurements for Alice, $\left| {A_s^k} \right\rangle $, and those for Bob, $\left| {B_t^k} \right\rangle $, are completely determined by ${A^1} = \{ \left| {A_1^1} \right\rangle ,\left| {A_2^1} \right\rangle , \ldots ,\left| {A_d^1} \right\rangle \} $, which corresponds to a SU($d$) unitary matrix, ${A^1} = {(a_{s,m}^1)_{d \times d}}$, with each element being the weight amplitude of the OAM eigenmodes. Hereafter we denote by $P^{\rm{opt}}$ the optimal value of Hardy fraction by ranging over all unitary matrices $A^1$ with a given optimal OAM Hardy state. In our experiment, we demonstrate the Hardy's paradox for the general $(k, d)$ systems with the optimal Hardy states, which are prepared from the original two-photon OAM entangled states and then are manipulated via the Procrustean method of entanglement concentration with two spatial light modulators (SLMs).

\section{III. EXPERIMENTAL SETUP AND RESULTS}

Our experimental setup is sketched in Fig. 1, which has been employed for demonstrating the angular or radial version of EPR paradox \cite{Leach10, Chen19}, based on the reconfigurable feature of SLMs. A mode-locked 355 nm ultraviolet laser pumps a 3-mm-thick $\beta$-barium borate crystal (BBO) and creates 710 nm frequency-degenerate photon pairs collinearly. We use a longpass filter (IF1) behind the crystal to block the pump beam, and then use a non-polarizing beam splitter (BS) to separate the signal and idler photons. In each of down-converted arms, a 4-$f$ telescope consisting of two lenses ($f1$ = 100 mm and $f2$ = 400 mm) is constructed to image the output facet of BBO crystal onto both SLMs (Hamamatsu, X10486-1). Each SLM is loaded with specially designed holographic gratings both for preparing the desired measurement OAM states and for performing the entanglement concentration. Then another telescope ($f3$ = 1000 mm and $f4$ = 2 mm) is used to reimage the plane of SLM onto the input facet of single-mode fiber (SMF), which is connected to a single-photon counter (Excelitas, SPCM-AQRH-14-FC). Besides, two bandpass filters (IF2) centered at 710 nm with 10 nm width are placed in front of the SMF to reduce the detection of noise photons. The outputs of both single-photon counters are connected to a coincidence circuit with a time resolution of 25ns.
\begin{figure*}[t]
\centerline{\includegraphics[width=1.8\columnwidth]{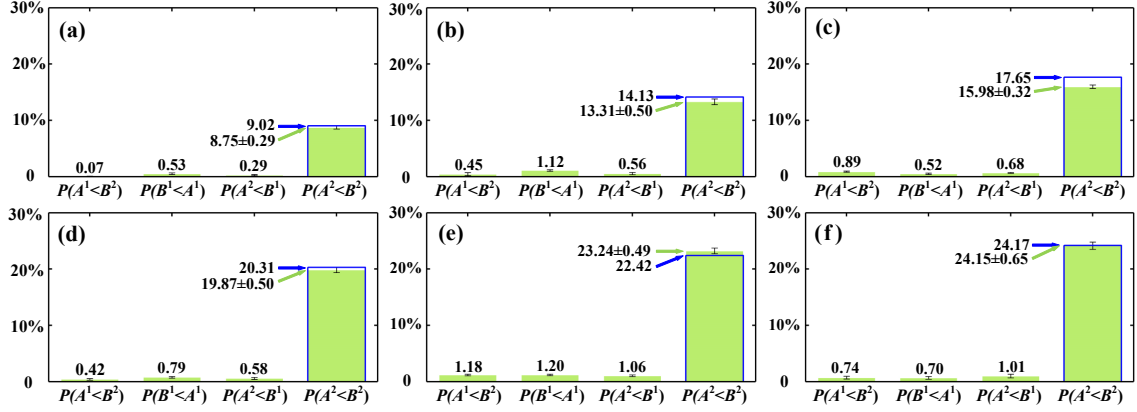}}
\caption{ (Color online) Hardy's paradox in $(k, 2)$ scenario: (a) $H_{\left( {2,2} \right)}^{\rm{opt}}$, (b) $H_{\left( {2,3} \right)}^{\rm{opt}}$, (c) $H_{\left( {2,4} \right)}^{\rm{opt}}$, (d) $H_{\left( {2,5} \right)}^{\rm{opt}}$, (e) $H_{\left( {2,6} \right)}^{\rm{opt}}$, and (f) $H_{\left( {2,7} \right)}^{\rm{opt}}$. The empty bars (blue edges) are the theoretical predictions while the solid bars (green) are experimental results.}
\end{figure*}

In our first set of experiment, we restrict our attention to two-setting Hardy's proof for high-dimensional OAM subspace, i.e., in $(2, d)$ scenario with the dimension ranging from $d$ = 2 to 7. We consider the original scheme proposed by Chen \emph{et al}. \cite{Chen13}. However, their optimal Hardy states are represented by upper triangular matrices, which obviously differ from the diagonal ones generated originally via SPDC \cite{Mair01}. Thus we need apply the Schmidt decomposition \cite{Nielsen00,Jaeger07} to transform the upper triangular matrices into the diagonal ones. Generally, a bipartite pure state $\left| \Phi  \right\rangle $ with any fixed orthonormal bases $\left| u \right\rangle $ and $\left| v \right\rangle $ for Alice and Bob can be expressed as, $\left| \Phi  \right\rangle  = \sum\nolimits_{uv} {{a_{uv}}\left| u \right\rangle \left| v \right\rangle } $, where ${a_{uv}}$ is complex number. We assume its Schmidt decomposition, $\left| \Phi  \right\rangle  = \sum\nolimits_g {{\lambda _g}\left| {{g_A}} \right\rangle } \left| {{g_B}} \right\rangle $, where $\left| {{g_A}} \right\rangle  = \sum\nolimits_u {{x_{ug}}\left| u \right\rangle } $ and $\left| {{g_B}} \right\rangle  = \sum\nolimits_v {{y_{gv}}\left| v \right\rangle } $ are the Schmidt bases for photon A and B, respectively, and ${\lambda _g}$ is the weight amplitude \cite{Nielsen00}. As $\left| {{g_A}} \right\rangle $ and $\left| {{g_B}} \right\rangle $ are two orthonormal bases in their own spaces, our experiment can naturally employ the OAM eigenstates to represent them. After some algebra, we obtain the diagonal matrices in $(2, d)$ scenario, as follows,
\begin{subequations}
\begin{align}
H_{\left( {2,2} \right)}^{\rm{opt}} = \rm{diag}&\left( {0.9070,{\rm{ }}0.4211} \right),\\
H_{\left( {2,3} \right)}^{\rm{opt}} = \rm{diag}&\left( 0.8585,{\rm{ }}0.4040,{\rm{ }}0.3159 \right),\\
H_{\left( {2,4} \right)}^{\rm{opt}} = \rm{diag}&\left( 0.8263,{\rm{ }}0.3947,{\rm{ }}0.3013,{\rm{ }}0.2657 \right),\\
H_{\left( {2,5} \right)}^{\rm{opt}} = \rm{diag}&\left( 0.8024,{\rm{ }}0.3882,{\rm{ }}0.2938,{\rm{ }}0.2532,{\rm{ }}0.2344 \right),\\
H_{\left( {2,6} \right)}^{\rm{opt}} = \rm{diag}&\left( 0.7837,{\rm{ }}0.3830,{\rm{ }}0.2889,{\rm{ }}0.2462,\right.\nonumber\\
&\left.0.2237,{\rm{ }}0.2123 \right),\\
H_{\left( {2,7} \right)}^{\rm{opt}} = \rm{diag}&\left( 0.7683,{\rm{ }}0.3787,{\rm{ }}0.2852,0.2417,{\rm{ }}\right.\nonumber\\
&\left.0.2173,{\rm{ }}0.2030,{\rm{ }}0.1955 \right).
\end{align}
\end{subequations}

Another issue here is that the original OAM entangled states produced via SPDC are not exactly the same as the optimal Hardy states. For example, we have the original state, $H_{\left( {2,3} \right)}^{\rm{SPDC}} = \rm{diag}(0.7975,0.5316,0.2853)$, residing in the subspace spanned by two-photon OAM bases, ${\left| 0 \right\rangle _A}{\left| 0 \right\rangle _B}$, ${\left| +1 \right\rangle _A}{\left| -1 \right\rangle _B}$, and ${\left| +2 \right\rangle _A}{\left| -2 \right\rangle _B}$. In an effort to achieve the maximal probability of nonlocal events, here we need apply the so-called Procrustean method of entanglement concentration \cite{Bennett96} to tailor the original state $H_{\left( {2,3} \right)}^{\rm{SPDC}}$ into the desired optimal one $H_{\left( {2,3} \right)}^{\rm{opt}}$. For this, we choose the local operation to change the weight amplitude of each OAM mode by altering the diffraction efficiencies of the blazed phase gratings \cite{Dada11}. For each weight amplitude of the OAM modes in the original two-photon state, if it is larger than that in the optimal Hardy state, then we decrease the contrast of blazed phase grating to obtain a lower diffraction efficiency for this OAM mode, and therefore the weight amplitudes of each OAM mode between the optimal Hardy states and the experimental one can be equalized. Then, based on the numeric strategy mentioned above with the optimal Hardy states of Eq. (7), we can calculate the desired OAM measurement states, $\left| {A_s^i} \right\rangle $ and $\left| {B_t^j} \right\rangle $, all of which are specified and listed in Appendix 1. By loading these OAM superposition states on SLM and record the coincidence counts accordingly, we obtain the experimental observations as shown in Fig. 2. We can see that the optimal successful probability can reach $P_{\left( {2,2} \right)}^{\rm{opt}} = 8.75 \pm 0.29\% $, $P_{\left( {2,3} \right)}^{\rm{opt}} = 13.31 \pm 0.50\% $, $P_{\left( {2,4} \right)}^{{\rm{opt}}} = 15.98 \pm 0.32\% $, $P_{\left( {2,5} \right)}^{\rm{opt}} = 19.87 \pm 0.50\% $, $P_{\left( {2,6} \right)}^{\rm{opt}} = 23.24 \pm 0.49\% $, and $P_{\left( {2,7} \right)}^{\rm{opt}} = 24.15 \pm 0.65\% $. One can see that these results show a good agreement with the quantum-mechanical predictions. Besides, our observations from $d$ = 2 to $d$ = 7 confirms that the nonlocal events can significantly increase with the dimension regardless of two-setting configuration only, which obviously surpasses the bound limited by the original version \cite{Rabelo12}.
\begin{figure}[b]
\centerline{\includegraphics[width=0.8\columnwidth]{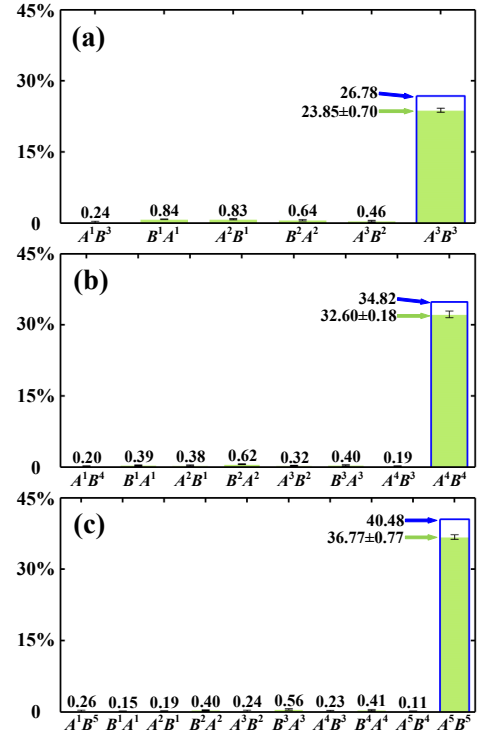}}
\caption{ (Color online) Hardy's paradox in $(k, 3)$ scenario: (a) $H_{\left( {3,3} \right)}^{\rm{opt}}$, (b) $H_{\left( {4,3} \right)}^{\rm{opt}}$, and (c) $H_{\left( {5,3} \right)}^{\rm{opt}}$. The empty bars (blue edges) are the theoretical values while the solid bars (green) are experimental results. ${A^i}{B^j}$ and ${B^j}{A^i}$ stand for $P\left( {{A^i} < {B^j}} \right)$ and $P\left( {{B^j} < {A^i}} \right)$, respectively.}
\end{figure}

In our second set of experiments, we further consider the general multisetting and multidimensional scenario $(k, d)$, where $k$ = 3, 4, 5 and $d$ = 3 in the OAM subspaces. Here we employ the diagonal matrices to represent the optimal Hardy states rather than that with the upper (or lower) triangle matrices in \cite{Meng18}. For this, we first apply the Schmidt decomposition to obtain the eigenvalues of their upper (or lower) triangle matrices and obtain that,
\begin{subequations}
\begin{align}
H_{\left( {3,3} \right)}^{\rm{opt}} = \rm{diag}\left( 0.8006,{\rm{ }}0.4578,{\rm{ }}0.3865 \right),\\
H_{\left( {4,3} \right)}^{\rm{opt}} = \rm{diag}\left( 0.7630,{\rm{ }}0.4856,{\rm{ }}0.4267 \right),\\
H_{\left( {5,3} \right)}^{\rm{opt}} = \rm{diag}\left( 0.7366,0.5025,0.4526 \right).
\end{align}
\end{subequations}

Then we look at the original entangled OAM spectrum (inset of Fig. 1), and exploit suitable OAM subspaces with entanglement concentration to prepare the optimal OAM Hardy states. After some similar algebra, we can calculate all the desired OAM measurements, which are listed in Appendix 2. As shown in Fig. 3, we experimentally obtain the successful probabilities, $P_{\left( {3,3} \right)}^{\rm{opt}} = 23.85 \pm 0.70\% $, $P_{\left( {4,3} \right)}^{\rm{opt}} = 32.60 \pm 0.18\% $, and $P_{\left( {5,3} \right)}^{\rm{opt}} = 36.77 \pm 0.77\% $, respectively, showing an excellent agreement with the theoretical predictions in \cite{Meng18}.

When we translate our theoretical findings to a real experiment, it is necessary to take the effect of imperfect measurements into account, as the observed correlation function generally deviates from that assumed in an ideal situation. Thus the derivation of a Bell-type inequality is practically desirable. In analogy to Refs. \cite{Boschi97,Karimi14} and in light of Eqs. (1)-(4), we introduce the Clauser-Horne inequality \cite{Mermin94},
\begin{eqnarray}
S_{\left( {k,d} \right)}=&P\left( {{A_k} < {B_k}} \right) - \sum\limits_{i = 2}^k {P\left( {{A_i} < {B_{i - 1}}} \right)} - \quad\, \nonumber\\
&\sum\limits_{i = 2}^k {P\left( {{B_{i - 1}} < {A_{i - 1}}} \right)}  - P\left( {{A_1} < {B_k}} \right) \le 0,
\end{eqnarray}
which is established in any local realistic theories. However, in quantum mechanics, all other probabilities can be zero except for the first terms, and therefore resulting in a violation of the inequality. For comparison, we present in Table 1 the measured $S$ values in $(k, 3)$ scenario for both the original states $H_{\left( {k,3} \right)}^{\rm{SPDC}}$ and the optimal states $H_{\left( {k,3} \right)}^{\rm{opt}}$. Our experimental results can violate the inequality by 58 standard deviations, thus confirming the contextual behavior of quantum mechanics.

\begin{table}[t]
\caption{Experimental ${S_{\left( {k,3} \right)}}$ for the original states $H_{\left( {k,3} \right)}^{\rm{SPDC}}$ and the optimal states $H_{\left( {k,3} \right)}^{\rm{opt}}$ in $(k, 3)$ scenario.}
\begin{tabular}{|c|c|c|c|}
\hline
\thead{ }   & \thead{$k$ = 3} & \thead{$k$ = 4} & \thead{$k$ = 5} \\
\hline
\thead{$S_{\left( {k,3} \right)}^{\rm{SPDC}}$ }   & \thead{$15.27 \pm 0.58\% $} & \thead{$23.58 \pm 0.60\% $} & \thead{$26.55 \pm 0.73\% $} \\
\hline
\thead{$S_{\left( {k,3} \right)}^{\rm{opt}}$ }  & \thead{$20.85 \pm 0.53\% $} & \thead{$29.76 \pm 0.75\% $} & \thead{$34.23 \pm 0.59\% $} \\
\hline
\end{tabular}
\end{table}

\section{IV. CONCLUSIONS}

In summary, we have presented the first experiment to demonstrate the Hardy's paradox for multisetting and multidimensional systems by exploiting two-photon entangled OAM states. The experimental results in $(2, d)$ scenario with the dimension up to $d$ = 7 revealed that the nonlocal events could increase with the dimension, significantly surpassing the analogue of Tsirelson’s bound in the original version. Besides, we have achieved a maximum successful probability up to 36.77\% for $(k, 3)$ scenario with the ladder up to $k$ = 5. Our experiment demonstrates evidently that both the multisetting and multidimensional features of the used quantum systems can yield a much sharper contradiction between quantum mechanics and classical theories. For some quantum communication tasks, our scheme of tailoring high-dimensional OAM entangled states may offer some other advantages, for example, non-classical correlations can be made more robust to the presence of noise and other deleterious environmental effects \cite{Erhard18,Ecker19}.

\section{ACKNOWLEDGMENTS}

We are grateful to Miles Padgett at the University of Glasgow for kind support. This work is supported by the National Natural Science Foundation of China (61975169, 91636109), the Fundamental Research Funds for the Central Universities at Xiamen University (20720190057, 20720190054), the Natural Science Foundation of Fujian Province of China for Distinguished Young Scientists (2015J06002), and the program for New Century Excellent Talents in University of China (NCET-13-0495).

\section{APPENDIX}
{\bf 1. The desired OAM measurement states for (2, $d$) with the dimension ranging from $d$ = 2 to 7.}

In our actual experiment, we choose the OAM modes ${\ell _1} = 0,{\ell _2} = +1$ for two-dimensional optimal state $H_{\left( {2,2} \right)}^{{\rm{opt}}} $; ${\ell _1} = 0,{\ell _2} = +1,{\ell _3} = +2$ for three-dimensional optimal state $H_{\left( {2,3} \right)}^{{\rm{opt}}} $; ${\ell _1} = 0,{\ell _2} = +1,{\ell _3} = +2,{\ell _4} = -2$ for four-dimensional optimal state  $H_{\left( {2,4} \right)}^{{\rm{opt}}} $; ${\ell _1} = 0,{\ell _2} = +1,{\ell _3} = -1,{\ell _4} = +2,{\ell _5} = -2$ for five-dimensional optimal state $H_{\left( {2,5} \right)}^{{\rm{opt}}} $; ${\ell _1} = 0,{\ell _2} = +1,{\ell _3} = +2,{\ell _4} = -2,{\ell _5} = +3,{\ell _6} = -3$ for six-dimensional optimal state $H_{\left( {2,6} \right)}^{{\rm{opt}}} $ and ${\ell _1} = 0,{\ell _2} = +1,{\ell _3} = -1,{\ell _4} = +2,{\ell _5} = -2,{\ell _6} = +3,{\ell _7} = -3$ for seven-dimensional optimal state $H_{\left( {2,7} \right)}^{{\rm{opt}}}$. The desired OAM measurement states, $\left| {A_s^i} \right\rangle$ and $\left| {B_t^j} \right\rangle $, in (2, 2) scenario are:

\begin{subequations}
\begin{align*}
\left| {{A_{1,1}}} \right\rangle  = \left[ \begin{array}{c}
0.8264\\0.5631\end{array} \right],
\left| {{A_{1,2}}} \right\rangle  = \left[ \begin{array}{c}
 - 0.5631\\0.8264\end{array} \right],\nonumber\\
\left| {{A_{2,1}}} \right\rangle  = \left[ \begin{array}{c}
0.3016\\0.9534\end{array} \right],
\left| {{A_{2,2}}} \right\rangle  = \left[ \begin{array}{c}
 - 0.9534\\0.3016\end{array} \right],\nonumber\\
\left| {{B_{1,1}}} \right\rangle  = \left[ \begin{array}{c}
0.5631\\0.8264\end{array} \right],
\left| {{B_{1,2}}} \right\rangle  = \left[ \begin{array}{c}
 - 0.8264\\0.5631\end{array} \right],\nonumber\\
\left| {{B_{2,1}}} \right\rangle  = \left[ \begin{array}{c}
0.9534\\0.3016\end{array} \right],
\left| {{B_{2,2}}} \right\rangle  = \left[ \begin{array}{c}
 - 0.3016\\0.9534\end{array} \right],\nonumber
\end{align*}
\end{subequations}

The desired OAM measurement states, $\left| {A_s^i} \right\rangle $ and $\left| {B_t^j} \right\rangle $, in (2, 3) scenario are:
\begin{subequations}
\begin{align*}
&\left| {{A_{1,1}}} \right\rangle  = \left[ \begin{array}{c}
 - 0.7232\\0.6129\\0.3182\end{array} \right],
\left| {{A_{1,2}}} \right\rangle  = \left[ \begin{array}{c}
0.5484\\0.2295\\0.8041\end{array} \right],\nonumber\\
&\left| {{A_{1,3}}} \right\rangle  = \left[ \begin{array}{c}
 - 0.4198\\ - 0.7561\\0.5021\end{array} \right],
\left| {{A_{2,1}}} \right\rangle  = \left[ \begin{array}{c}
 - 0.1953\\0.7475\\0.6349\end{array} \right],\nonumber\\
&\left| {{A_{2,2}}} \right\rangle  = \left[ \begin{array}{c}
 - 0.3437\\0.5541\\ - 0.7582\end{array} \right],
\left| {{A_{2,3}}} \right\rangle  = \left[ \begin{array}{c}
 - 0.9185\\ - 0.3663\\0.1487\end{array} \right],\nonumber\\
&\left| {{B_{1,1}}} \right\rangle  = \left[ \begin{array}{c}
 - 0.4198\\0.7561\\0.5021\end{array} \right],
\left| {{B_{1,2}}} \right\rangle  = \left[ \begin{array}{c}
 - 0.5484\\0.2295\\ - 0.8041\end{array} \right],\nonumber\\
&\left| {{B_{1,3}}} \right\rangle  = \left[ \begin{array}{c}
 - 0.7232\\ - 0.6129\\0.3182\end{array} \right],
\left| {{B_{2,1}}} \right\rangle  = \left[ \begin{array}{c}
 - 0.9185\\0.3663\\0.1487\end{array} \right],\nonumber\\
&\left| {{B_{2,2}}} \right\rangle  = \left[ \begin{array}{l}
0.3437\\0.5541\\0.7582\end{array} \right],
\left| {{B_{2,3}}} \right\rangle  = \left[ \begin{array}{c}
 - 0.1953\\ - 0.7475\\0.6349\end{array} \right].\nonumber
\end{align*}
\end{subequations}
The desired OAM measurement states, $\left| {A_s^i} \right\rangle $ and  $\left| {B_t^j} \right\rangle $, in (2, 4) scenario are:
\begin{subequations}
\begin{align*}
&\left| {{A_{1,1}}} \right\rangle  = \left[ \begin{array}{c}
 - 0.6536\\0.6006\\ - 0.4108\\ - 0.2083\end{array} \right],
\left| {{A_{1,2}}} \right\rangle  = \left[ \begin{array}{c}
 - 0.5128\\0.0602\\0.5883\\0.6224\end{array} \right],\nonumber\\
&\left| {{A_{1,3}}} \right\rangle  = \left[ \begin{array}{c}
 - 0.4407\\ - 0.4562\\0.3767\\ - 0.6751\end{array} \right],
\left| {{A_{1,4}}} \right\rangle  = \left[ \begin{array}{c}
 - 0.3400\\ - 0.6539\\ - 0.5859\\0.3369\end{array} \right] ,\nonumber\\
&\left| {{A_{2,1}}} \right\rangle  = \left[ \begin{array}{c}
 - 0.1428\\0.5750\\ - 0.6749\\ - 0.4400\end{array} \right],
\left| {{A_{2,2}}} \right\rangle  = \left[ \begin{array}{c}
 - 0.2480\\0.6508\\0.1493\\0.7019\end{array} \right],\nonumber\\
&\left| {{A_{2,3}}} \right\rangle  = \left[ \begin{array}{c}
 - 0.3490\\0.3040\\0.6931\\ - 0.5526\end{array} \right],
\left| {{A_{2,4}}} \right\rangle  = \left[ \begin{array}{c}
 - 0.8924\\ - 0.3917\\ - 0.2045\\0.0915\end{array} \right],\nonumber\\
&\left| {{B_{1,1}}} \right\rangle  = \left[ \begin{array}{c}
 - 0.3400\\0.6539\\ - 0.5859\\ - 0.3369\end{array} \right],
\left| {{B_{1,2}}} \right\rangle  = \left[ \begin{array}{c}
 - 0.4407\\0.4562\\0.3767\\0.6751\end{array} \right],\nonumber\\
&\left| {{B_{1,3}}} \right\rangle  = \left[ \begin{array}{c}
 - 0.5128\\ - 0.0602\\0.5883\\ - 0.6224\end{array} \right],
\left| {{B_{1,4}}} \right\rangle  = \left[ \begin{array}{c}
 - 0.6536\\ - 0.6006\\ - 0.4108\\0.2083\end{array} \right],\nonumber\\
&\left| {{B_{2,1}}} \right\rangle  = \left[ \begin{array}{c}
 - 0.8924\\0.3917\\ - 0.2045\\ - 0.0915\end{array} \right],
\left| {{B_{2,2}}} \right\rangle  = \left[ \begin{array}{c}
 - 0.3490\\ - 0.3040\\0.6931\\0.5526\end{array} \right],\nonumber\\
&\left| {{B_{2,3}}} \right\rangle  = \left[ \begin{array}{c}
 - 0.2480\\ - 0.6508\\0.1493\\ - 0.7019\end{array} \right],
\left| {{B_{2,4}}} \right\rangle  = \left[ \begin{array}{c}
 - 0.1428\\ - 0.5750\\ - 0.6749\\0.4400\end{array} \right].\nonumber
\end{align*}
\end{subequations}
The desired OAM measurement states, $\left| {A_s^i} \right\rangle $ and  $\left| {B_t^j} \right\rangle $, in (2, 5) scenario are:
\begin{subequations}
\begin{align*}
&\left| {{A_{1,1}}} \right\rangle  = \left[ \begin{array}{c}
 - 0.6023\\0.5774\\0.4400\\ - 0.2965\\0.1493\end{array} \right],
\left| {{A_{1,2}}} \right\rangle  = \left[ \begin{array}{c}
 - 0.4799\\0.1963\\ - 0.3300\\0.6254\\ - 0.4808\end{array} \right],\nonumber\\
&\left| {{A_{1,3}}} \right\rangle  = \left[ \begin{array}{c}
  - 0.4304\\ - 0.2028\\ - 0.6054\\ - 0.1019\\0.6299\end{array} \right],
\left| {{A_{1,4}}} \right\rangle  = \left[ \begin{array}{c}
 - 0.3725\\ - 0.5111\\ - 0.0308\\ - 0.5558\\ - 0.5387\end{array} \right],\nonumber\\
&\left| {{A_{1,5}}} \right\rangle  = \left[ \begin{array}{c}
 - 0.2879\\ - 0.5707\\0.5745\\0.4491\\0.2443\end{array} \right],
\left| {{A_{2,1}}} \right\rangle  = \left[ \begin{array}{c}
0.1116\\ - 0.4570\\ - 0.6080\\0.5514\\ - 0.3240\end{array} \right],\nonumber\\
&\left| {{A_{2,2}}} \right\rangle  = \left[ \begin{array}{c}
 - 0.1931\\0.6078\\0.2270\\0.4517\\ - 0.5811\end{array} \right],
\left| {{A_{2,3}}} \right\rangle  = \left[ \begin{array}{c}
 - 0.2667\\0.4821\\ - 0.4890\\0.2769\\0.6170\end{array} \right],\nonumber\\
&\left| {{A_{2,4}}} \right\rangle  = \left[ \begin{array}{c}
 - 0.3453\\0.1607\\ - 0.5342\\ - 0.6300\\ - 0.4155\end{array} \right],
\left| {{A_{2,5}}} \right\rangle  = \left[ \begin{array}{c}
 - 0.8717\\ - 0.4043\\0.2331\\0.1354\\0.0631\end{array} \right],\nonumber\\
&\left| {{B_{1,1}}} \right\rangle  = \left[ \begin{array}{c}
 - 0.2879\\0.5707\\0.5745\\ - 0.4491\\0.2443\end{array} \right],
\left| {{B_{1,2}}} \right\rangle  = \left[ \begin{array}{c}
 - 0.3725\\0.5111\\ - 0.0308\\0.5558\\ - 0.5387\end{array} \right],\nonumber\\
&\left| {{B_{1,3}}} \right\rangle  = \left[ \begin{array}{c}
 - 0.4304\\0.2028\\ - 0.6054\\0.1019\\0.6299\end{array} \right],
\left| {{B_{1,4}}} \right\rangle  = \left[ \begin{array}{c}
 - 0.4799\\ - 0.1963\\ - 0.3300\\ - 0.6254\\ - 0.4808\end{array} \right],\nonumber\\
&\left| {{B_{1,5}}} \right\rangle  = \left[ \begin{array}{c}
 - 0.6023\\ - 0.5774\\0.4400\\0.2965\\0.1493\end{array} \right],
\left| {{B_{2,1}}} \right\rangle  = \left[ \begin{array}{c}
 - 0.8717\\0.4043\\0.2331\\ - 0.1354\\0.0631\end{array} \right],\nonumber\\
&\left| {{B_{2,2}}} \right\rangle  = \left[ \begin{array}{c}
 - 0.3453\\ - 0.1607\\ - 0.5342\\0.6300\\ - 0.4155\end{array} \right],
\left| {{B_{2,3}}} \right\rangle  = \left[ \begin{array}{c}
 - 0.2667\\ - 0.4821\\ - 0.4890\\ - 0.2769\\0.6170\end{array} \right],\nonumber\\
&\left| {{B_{2,4}}} \right\rangle  = \left[ \begin{array}{c}
 - 0.1931\\ - 0.6078\\0.2270\\ - 0.4517\\ - 0.5811\end{array} \right],
\left| {{B_{2,5}}} \right\rangle  = \left[ \begin{array}{c}
0.1116\\0.4570\\ - 0.6080\\ - 0.5514\\ - 0.3240\end{array} \right].\nonumber
\end{align*}
\end{subequations}
The desired OAM measurement states, $\left| {A_s^i} \right\rangle $ and  $\left| {B_t^j} \right\rangle $, in (2, 6) scenario are:
\begin{subequations}
\begin{align*}
&\left| {{A_{1,1}}} \right\rangle  = \left[ \begin{array}{c}
 - 0.5624\\0.5536\\0.4466\\ - 0.3372\\0.2262\\0.1135\end{array} \right],
\left| {{A_{1,2}}} \right\rangle  = \left[ \begin{array}{c}
 - 0.4517\\0.2639\\ - 0.1457\\0.4810\\ - 0.5734\\ - 0.3807\end{array} \right],\nonumber\\
&\left| {{A_{1,3}}} \right\rangle  = \left[ \begin{array}{c}
 - 0.4139\\ - 0.0385\\ - 0.5508\\0.3179\\0.3520\\0.5467\end{array} \right],
\left| {{A_{1,4}}} \right\rangle  = \left[ \begin{array}{c}
 - 0.3744\\ - 0.3271\\ - 0.3943\\ - 0.4788\\0.2186\\ - 0.5659\end{array} \right],\nonumber\\
&\left| {{A_{1,5}}} \right\rangle  = \left[ \begin{array}{c}
 - 0.3243\\ - 0.5098\\0.1682\\ - 0.3088\\ - 0.5686\\0.4336\end{array} \right],
\left| {{A_{1,6}}} \right\rangle  = \left[ \begin{array}{c}
 - 0.2510\\ - 0.5055\\0.5406\\0.4788\\0.3535\\ - 0.1870\end{array} \right],\nonumber\\
&\left| {{A_{2,1}}} \right\rangle  = \left[ \begin{array}{c}
0.0909\\ - 0.3747\\ - 0.5314\\0.5522\\ - 0.4488\\ - 0.2502\end{array} \right],
\left| {{A_{2,2}}} \right\rangle  = \left[ \begin{array}{c}
 - 0.1572\\0.5409\\0.3939\\0.1219\\ - 0.5343\\ - 0.4766\end{array} \right],\nonumber\\
&\left| {{A_{2,3}}} \right\rangle  = \left[ \begin{array}{c}
 - 0.2160\\0.5224\\ - 0.1655\\0.5604\\0.0844\\0.5761\end{array} \right],
\left| {{A_{2,4}}} \right\rangle  = \left[ \begin{array}{c}
 - 0.2733\\0.3458\\ - 0.5595\\ - 0.1260\\0.4536\\ - 0.5207\end{array} \right],\nonumber\\
&\left| {{A_{2,5}}} \right\rangle  = \left[ \begin{array}{c}
 - 0.3392\\0.0737\\ - 0.3993\\ - 0.5695\\ - 0.5391\\0.3242\end{array} \right],
\left| {{A_{2,6}}} \right\rangle  = \left[ \begin{array}{c}
 - 0.8548\\ - 0.4112\\0.2502\\0.1610\\0.0981\\ - 0.0467\end{array} \right],\nonumber\\
&\left| {{B_{1,1}}} \right\rangle  = \left[ \begin{array}{c}
 - 0.2510\\0.5055\\0.5406\\ - 0.4788\\0.3535\\0.1870\end{array} \right],
\left| {{B_{1,2}}} \right\rangle  = \left[ \begin{array}{c}
 - 0.3243\\0.5098\\0.1682\\0.3088\\ - 0.5686\\ - 0.4336\end{array} \right],\nonumber\\
&\left| {{B_{1,3}}} \right\rangle  = \left[ \begin{array}{c}
 - 0.3744\\0.3271\\ - 0.3943\\0.4788\\0.2186\\0.5659\end{array} \right],
\left| {{B_{1,4}}} \right\rangle  = \left[ \begin{array}{c}
 - 0.4139\\0.0385\\ - 0.5508\\ - 0.3179\\0.3520\\ - 0.5467\end{array} \right],\nonumber\\
&\left| {{B_{1,5}}} \right\rangle  = \left[ \begin{array}{c}
 - 0.4517\\ - 0.2639\\ - 0.1457\\ - 0.4810\\ - 0.5734\\0.3807\end{array} \right],
\left| {{B_{1,6}}} \right\rangle  = \left[ \begin{array}{c}
 - 0.5624\\ - 0.5536\\0.4466\\0.3372\\0.2262\\ - 0.1135\end{array} \right],\nonumber\\
&\left| {{B_{2,1}}} \right\rangle  = \left[ \begin{array}{c}
 - 0.8548\\0.4112\\0.2502\\ - 0.1610\\0.0981\\0.0467\end{array} \right],
\left| {{B_{2,2}}} \right\rangle  = \left[ \begin{array}{c}
 - 0.3392\\ - 0.0737\\ - 0.3993\\0.5695\\ - 0.5391\\ - 0.3242\end{array} \right],\nonumber\\
&\left| {{B_{2,3}}} \right\rangle  = \left[ \begin{array}{c}
 - 0.2733\\ - 0.3458\\ - 0.5595\\0.1260\\0.4536\\0.5207\end{array} \right],
\left| {{B_{2,4}}} \right\rangle  = \left[ \begin{array}{c}
 - 0.2160\\ - 0.5224\\ - 0.1655\\ - 0.5604\\0.0844\\ - 0.5761\end{array} \right],\nonumber\\
&\left| {{B_{2,5}}} \right\rangle  = \left[ \begin{array}{c}
 - 0.1572\\ - 0.5409\\0.3939\\ - 0.1219\\ - 0.5343\\0.4766\end{array} \right],
\left| {{B_{2,6}}} \right\rangle  = \left[ \begin{array}{c}
0.0909\\0.3747\\ - 0.5314\\ - 0.5522\\ - 0.4488\\0.2502\end{array} \right].\nonumber
\end{align*}
\end{subequations}
The desired OAM measurement states, $\left| {A_s^i} \right\rangle $ and  $\left| {B_t^j} \right\rangle $, in (2, 7) scenario are:
\begin{subequations}
\begin{align*}
&\left| {{A_{1,1}}} \right\rangle  = \left[ \begin{array}{c}
 - 0.5302\\0.5315\\0.4440\\ - 0.3562\\0.2683\\ - 0.1796\\0.0901\end{array} \right] ,
\left| {{A_{1,2}}} \right\rangle  = \left[ \begin{array}{c}
 - 0.4277\\0.2985\\ - 0.0224\\0.3327\\ - 0.5152\\0.5055\\ - 0.3095\end{array} \right],\nonumber\\
&\left| {{A_{1,3}}} \right\rangle  = \left[ \begin{array}{c}
 - 0.3969\\0.0658\\ - 0.4281\\0.4794\\ - 0.0298\\ - 0.4529\\0.4681\end{array} \right],
\left| {{A_{1,4}}} \right\rangle  = \left[ \begin{array}{c}
 - 0.3674\\ - 0.1796\\ - 0.5044\\ - 0.1196\\0.5251\\0.0602\\ - 0.5335\end{array} \right],\nonumber\\
&\left| {{A_{1,5}}} \right\rangle  = \left[ \begin{array}{c}
 - 0.3327\\ - 0.3824\\ - 0.1875\\ - 0.5294\\ - 0.2060\\0.3765\\0.4935\end{array} \right],
\left| {{A_{1,6}}} \right\rangle  = \left[ \begin{array}{c}
 - 0.2881\\ - 0.4895\\0.2757\\ - 0.1041\\ - 0.4289\\ - 0.5290\\ - 0.3559\end{array} \right],\nonumber\\
&\left| {{A_{1,7}}} \right\rangle  = \left[ \begin{array}{c}
 - 0.2232\\ - 0.4539\\0.5034\\0.4767\\0.3993\\0.2861\\0.1490\end{array} \right],
\left| {{A_{2,1}}} \right\rangle  = \left[ \begin{array}{c}
0.0764\\ - 0.3151\\ - 0.4639\\0.5185\\ - 0.4832\\0.3704\\ - 0.2004\end{array} \right],\nonumber\\
&\left| {{A_{2,2}}} \right\rangle  = \left[ \begin{array}{c}
 - 0.1319\\0.4771\\0.4547\\ - 0.1086\\ - 0.3147\\0.5301\\ - 0.3954\end{array} \right],
\left| {{A_{2,3}}} \right\rangle  = \left[ \begin{array}{c}
 - 0.1810\\0.5105\\0.0721\\0.4840\\ - 0.3466\\ - 0.2906\\0.5125\end{array} \right],\nonumber\\
&\left| {{A_{2,4}}} \right\rangle  = \left[ \begin{array}{c}
 - 0.2274\\0.4240\\ - 0.3691\\0.3183\\0.4726\\ - 0.1696\\ - 0.5280\end{array} \right],
\left| {{A_{2,5}}} \right\rangle  = \left[ \begin{array}{c}
 - 0.2747\\0.2450\\ - 0.5291\\ - 0.3454\\0.1384\\0.5037\\0.4387\end{array} \right],\nonumber\\
&\left| {{A_{2,6}}} \right\rangle  = \left[ \begin{array}{c}
 - 0.3325\\0.0177\\ - 0.2972\\ - 0.4826\\ - 0.5389\\ - 0.4574\\ - 0.2613\end{array} \right],
\left| {{A_{2,7}}} \right\rangle  = \left[ \begin{array}{c}
 - 0.8404\\ - 0.4152\\0.2613\\0.1776\\0.1203\\0.0752\\0.0363\end{array} \right],\nonumber\\
&\left| {{B_{1,1}}} \right\rangle  = \left[ \begin{array}{c}
 - 0.2232\\0.4539\\0.5034\\ - 0.4767\\0.3993\\ - 0.2861\\0.1490\end{array} \right],
\left| {{B_{1,2}}} \right\rangle  = \left[ \begin{array}{c}
 - 0.2881\\0.4895\\0.2757\\0.1041\\ - 0.4289\\0.5290\\ - 0.3559\end{array} \right],\nonumber\\
&\left| {{B_{1,3}}} \right\rangle  = \left[ \begin{array}{c}
 - 0.3327\\0.3824\\ - 0.1875\\0.5294\\ - 0.2060\\ - 0.3765\\0.4935\end{array} \right],
\left| {{B_{1,4}}} \right\rangle  = \left[ \begin{array}{c}
 - 0.3674\\0.1796\\ - 0.5044\\0.1196\\0.5251\\ - 0.0602\\ - 0.5335\end{array} \right],\nonumber\\
&\left| {{B_{1,5}}} \right\rangle  = \left[ \begin{array}{c}
 - 0.3969\\ - 0.0658\\ - 0.4281\\ - 0.4794\\ - 0.0298\\0.4529\\0.4681\end{array} \right],
\left| {{B_{1,6}}} \right\rangle  = \left[ \begin{array}{c}
 - 0.4277\\ - 0.2985\\ - 0.0224\\ - 0.3327\\ - 0.5152\\ - 0.5055\\ - 0.3095\end{array} \right],\nonumber\\
&\left| {{B_{1,7}}} \right\rangle  = \left[ \begin{array}{c}
 - 0.5302\\ - 0.5315\\0.4440\\0.3562\\0.2683\\0.1796\\0.0901\end{array} \right],
\left| {{B_{2,1}}} \right\rangle  = \left[ \begin{array}{c}
 - 0.8404\\0.4152\\0.2613\\ - 0.1776\\0.1203\\ - 0.0752\\0.0363\end{array} \right],\nonumber\\
&\left| {{B_{2,2}}} \right\rangle  = \left[ \begin{array}{c}
 - 0.3325\\ - 0.0177\\ - 0.2972\\0.4826\\ - 0.5389\\0.4574\\ - 0.2613\end{array} \right],
\left| {{B_{2,3}}} \right\rangle  = \left[ \begin{array}{c}
 - 0.2747\\ - 0.2450\\ - 0.5291\\0.3454\\0.1384\\  - 0.5037\\ 0.4387\end{array} \right],\nonumber\\
&\left| {{B_{2,4}}} \right\rangle  = \left[ \begin{array}{c}
 - 0.2274\\ - 0.4240\\ - 0.3691\\ - 0.3183\\0.4726\\0.1696\\ - 0.5280\end{array} \right],
\left| {{B_{2,5}}} \right\rangle  = \left[ \begin{array}{c}
 - 0.1810\\ - 0.5105\\.0721\\ - 0.4840\\ - 0.3466\\0.2906\\0.5125\end{array} \right],\nonumber\\
&\left| {{B_{2,6}}} \right\rangle  = \left[ \begin{array}{c}
 - 0.1319\\ - 0.4771\\0.4547\\0.1086\\ - 0.3147\\ - 0.5301\\ - 0.3954\end{array} \right],
\left| {{B_{2,7}}} \right\rangle  = \left[ \begin{array}{c}
0.0764\\0.3151\\ - 0.4639\\ - 0.5185\\ - 0.4832\\ - 0.3704\\ - 0.2004\end{array} \right].\nonumber
\end{align*}
\end{subequations}

{\bf 2. The desired OAM measurement states for ($k$, 3) scenario with $k$ = 3, 4, 5}

 In second experiment, we choose the OAM modes ${\ell _1} = 0,{\ell _2} = +1,{\ell _3} = +2$ for optimal states $H_{\left( {3,3} \right)}^{{\rm{opt}}}$ and $H_{\left( {4,3} \right)}^{{\rm{opt}}}$; ${\ell _1} = 0,{\ell _2} = +1,{\ell _3} = -1$ for optimal states $H_{\left( {5,3} \right)}^{{\rm{opt}}}$. The desired OAM measurement states, $\left| {A_s^i} \right\rangle $ and  $\left| {B_t^j} \right\rangle $, in (3, 3) scenario are:
\begin{subequations}
\begin{align*}
&\left| {{A_{1,1}}} \right\rangle  = \left[ \begin{array}{c}
 - 0.8625\\0.4606\\0.2097\end{array} \right],
\left| {{A_{1,2}}} \right\rangle  = \left[ \begin{array}{c}
0.4287\\0.4448\\0.7863\end{array} \right],\nonumber\\
&\left| {{A_{1,3}}} \right\rangle  = \left[ \begin{array}{c}
 - 0.2689\\ - 0.7681\\0.5811\end{array} \right],
\left| {{A_{2,1}}} \right\rangle  = \left[ \begin{array}{c}
 - 0.4585\\0.7489\\0.4784\end{array} \right],\nonumber\\
&\left| {{A_{2,2}}} \right\rangle  = \left[ \begin{array}{c}
 - 0.5651\\0.1697\\ - 0.8074\end{array} \right],
\left| {{A_{2,3}}} \right\rangle  = \left[ \begin{array}{c}
 - 0.6858\\ - 0.6406\\0.3454\end{array} \right],\nonumber\\
&\left| {{A_{3,1}}} \right\rangle  = \left[ \begin{array}{c}
 - 0.1475\\0.7365\\0.6601\end{array} \right],
\left| {{A_{3,2}}} \right\rangle  = \left[ \begin{array}{c}
0.2738\\ - 0.611\\0.7428\end{array} \right],\nonumber\\
&\left| {{A_{3,3}}} \right\rangle  = \left[ \begin{array}{c}
 - 0.9504\\ - 0.2903\\0.1115\end{array} \right],
\left| {{B_{1,1}}} \right\rangle  = \left[ \begin{array}{c}
 - 0.6858\\0.6406\\0.3454\end{array} \right],\nonumber\\
&\left| {{B_{1,2}}} \right\rangle  = \left[ \begin{array}{c}
0.5651\\0.1697\\0.8074\end{array} \right],
\left| {{B_{1,3}}} \right\rangle  = \left[ \begin{array}{c}
 - 0.4585\\ - 0.7489\\0.4784\end{array} \right],\nonumber\\
&\left| {{B_{2,1}}} \right\rangle  = \left[ \begin{array}{c}
0.2689\\ - 0.7681\\ - 0.5811\end{array} \right],
\left| {{B_{2,2}}} \right\rangle  = \left[ \begin{array}{c}
 - 0.4287\\0.4448\\ - 0.7863\end{array} \right],\nonumber\\
&\left| {{B_{2,3}}} \right\rangle  = \left[ \begin{array}{c}
 - 0.8625\\ - 0.4606\\0.2097\end{array} \right],
\left| {{B_{3,1}}} \right\rangle  = \left[ \begin{array}{c}
 - 0.9504\\0.2903\\0.1115\end{array} \right],\nonumber\\
&\left| {{B_{3,2}}} \right\rangle  = \left[ \begin{array}{c}
0.2738\\0.6110\\0.7428\end{array} \right],
\left| {{B_{3,3}}} \right\rangle  = \left[ \begin{array}{c}
0.1475\\0.7365\\ - 0.6601\end{array} \right].\nonumber
\end{align*}
\end{subequations}
The desired OAM measurement states, $\left| {A_s^i} \right\rangle $ and  $\left| {B_t^j} \right\rangle $, in (4, 3) scenario are:
\begin{subequations}
\begin{align*}
&\left| {{A_{1,1}}} \right\rangle  = \left[ \begin{array}{c}
 - 0.9176\\ - 0.366\\0.1553\end{array} \right],
\left| {{A_{1,2}}} \right\rangle  = \left[ \begin{array}{c}
 - 0.3440\\0.5348\\ - 0.7718\end{array} \right],\nonumber\\
&\left| {{A_{1,3}}} \right\rangle  = \left[ \begin{array}{c}
0.1994\\ - 0.7616\\ - 0.6166\end{array} \right],
\left| {{A_{2,1}}} \right\rangle  = \left[ \begin{array}{c}
 - 0.6648\\ - 0.6547\\0.3597\end{array} \right],\nonumber\\
&\left| {{A_{2,2}}} \right\rangle  = \left[ \begin{array}{c}
0.5718\\ - 0.1362\\0.8090\end{array} \right],
\left| {{A_{2,3}}} \right\rangle  = \left[ \begin{array}{c}
 - 0.4806\\0.7435\\0.4649\end{array} \right],\nonumber\\
&\left| {{A_{3,1}}} \right\rangle  = \left[ \begin{array}{c}
 - 0.3178\\ - 0.7725\\0.5497\end{array} \right],
\left| {{A_{3,2}}} \right\rangle  = \left[ \begin{array}{l}
0.4747\\0.3722\\0.7976\end{array} \right],\nonumber\\
&\left| {{A_{3,3}}} \right\rangle  = \left[ \begin{array}{c}
 - 0.8208\\0.5144\\0.2484\end{array} \right],
\left| {{A_{4,1}}} \right\rangle  = \left[ \begin{array}{c}
 - 0.1216\\ - 0.7300\\0.6726\end{array} \right],\nonumber\\
&\left| {{A_{4,2}}} \right\rangle  = \left[ \begin{array}{l}
0.2315\\0.6380\\0.7344\end{array} \right],
\left| {{A_{4,3}}} \right\rangle  = \left[ \begin{array}{c}
 - 0.9652\\0.2450\\0.0914\end{array} \right],\nonumber\\
&\left| {{B_{1,1}}} \right\rangle  = \left[ \begin{array}{c}
 - 0.8208\\ - 0.5144\\0.2484\end{array} \right],
\left| {{B_{1,2}}} \right\rangle  = \left[ \begin{array}{c}
0.4747\\ - 0.3722\\0.7976\end{array} \right],\nonumber\\
&\left| {{B_{1,3}}} \right\rangle  = \left[ \begin{array}{c}
0.3178\\ - 0.7725\\ - 0.5497\end{array} \right],
\left| {{B_{2,1}}} \right\rangle  = \left[ \begin{array}{c}
 - 0.4806\\ - 0.7435\\0.4649\end{array} \right],\nonumber\\
&\left| {{B_{2,2}}} \right\rangle  = \left[ \begin{array}{c}
0.5718\\0.1362\\0.8090\end{array} \right],
\left| {{B_{2,3}}} \right\rangle  = \left[ \begin{array}{c}
 - 0.6648\\0.6547\\0.3597\end{array} \right],\nonumber\\
&\left| {{B_{3,1}}} \right\rangle  = \left[ \begin{array}{c}
0.1994\\0.7616\\ - 0.6166\end{array} \right],
\left| {{B_{3,2}}} \right\rangle  = \left[ \begin{array}{c}
0.3440\\0.5348\\0.7718\end{array} \right],\nonumber\\
&\left| {{B_{3,3}}} \right\rangle  = \left[ \begin{array}{c}
 - 0.9176\\0.3660\\0.1553\end{array} \right],
\left| {{B_{4,1}}} \right\rangle  = \left[ \begin{array}{c}
 - 0.9652\\ - 0.245\\0.0914\end{array} \right],\nonumber\\
&\left| {{B_{4,2}}} \right\rangle  = \left[ \begin{array}{c}
 - 0.2315\\0.6380\\ - 0.7344\end{array} \right],
\left| {{B_{4,3}}} \right\rangle  = \left[ \begin{array}{c}
0.1216\\ - 0.7300\\ - 0.6726\end{array} \right].\nonumber
\end{align*}
\end{subequations}
The desired OAM measurement states, $\left| {A_s^i} \right\rangle $ and  $\left| {B_t^j} \right\rangle $, in (5, 3) scenario are:
\begin{subequations}
\begin{align*}
&\left| {{A_{1,1}}} \right\rangle  = \left[ \begin{array}{c}
0.9441\\0.3053\\0.1240\end{array} \right],
\left| {{A_{1,2}}} \right\rangle  = \left[ \begin{array}{c}
 - 0.2879\\0.5812\\0.7612\end{array} \right],\nonumber\\
&\left| {{A_{1,3}}} \right\rangle  = \left[ \begin{array}{c}
 - 0.1603\\0.7543\\ - 0.6366\end{array} \right],
\left| {{A_{2,1}}} \right\rangle  = \left[ \begin{array}{c}
0.7896\\0.5487\\0.2747\end{array} \right],\nonumber\\
&\left| {{A_{2,2}}} \right\rangle  = \left[ \begin{array}{c}
0.5022\\ - 0.3206\\ - 0.8031\end{array} \right],
\left| {{A_{2,3}}} \right\rangle  = \left[ \begin{array}{c}
 - 0.3527\\0.7721\\ - 0.5287\end{array} \right],\nonumber\\
&\left| {{A_{3,1}}} \right\rangle  = \left[ \begin{array}{c}
0.4951\\0.7395\\0.4561\end{array} \right],
\left| {{A_{3,2}}} \right\rangle  = \left[ \begin{array}{c}
0.5752\\0.1145\\ - 0.8100\end{array} \right],\nonumber\\
&\left| {{A_{3,3}}} \right\rangle  = \left[ \begin{array}{c}
0.6512\\ - 0.6634\\0.3686\end{array} \right],
\left| {{A_{4,1}}} \right\rangle  = \left[ \begin{array}{c}
0.2408\\0.7727\\0.5874\end{array} \right],\nonumber\\
&\left| {{A_{4,2}}} \right\rangle  = \left[ \begin{array}{c}
0.3936\\0.4755\\ - 0.7868\end{array} \right],
\left| {{A_{4,3}}} \right\rangle  = \left[ \begin{array}{c}
0.8872\\ - 0.4206\\0.1896\end{array} \right],\nonumber\\
&\left| {{A_{5,1}}} \right\rangle  = \left[ \begin{array}{c}
0.1052\\0.7257\\0.6799\end{array} \right],
\left| {{A_{5,2}}} \right\rangle  = \left[ \begin{array}{c}
 - 0.2030\\ - 0.6536\\0.7291\end{array} \right],\nonumber\\
&\left| {{A_{5,3}}} \right\rangle  = \left[ \begin{array}{c}
0.9735\\ - 0.2148\\0.0786\end{array} \right],
\left| {{B_{1,1}}} \right\rangle  = \left[ \begin{array}{c}
0.8872\\0.4206\\0.1896\end{array} \right],\nonumber\\
&\left| {{B_{1,2}}} \right\rangle  = \left[ \begin{array}{c}
 - 0.3936\\0.4755\\0.7868\end{array} \right],
\left| {{B_{1,3}}} \right\rangle  = \left[ \begin{array}{c}
 - 0.2408\\0.7727\\ - 0.5874\end{array} \right],\nonumber\\
&\left| {{B_{2,1}}} \right\rangle  = \left[ \begin{array}{c}
0.6512\\0.6634\\0.3686\end{array} \right],
\left| {{B_{2,2}}} \right\rangle  = \left[ \begin{array}{c}
0.5752\\ - 0.1145\\ - 0.8100\end{array} \right],\nonumber\\
&\left| {{B_{2,3}}} \right\rangle  = \left[ \begin{array}{c}
0.4951\\ - 0.7395\\0.4561\end{array} \right],
\left| {{B_{3,1}}} \right\rangle  = \left[ \begin{array}{c}
0.3527\\0.7721\\0.5287\end{array} \right],\nonumber\\
&\left| {{B_{3,2}}} \right\rangle  = \left[ \begin{array}{c}
0.5022\\0.3206\\ - 0.8031\end{array} \right],
\left| {{B_{3,3}}} \right\rangle  = \left[ \begin{array}{c}
0.7896\\ - 0.5487\\0.2747\end{array} \right],\nonumber\\
&\left| {{B_{4,1}}} \right\rangle  = \left[ \begin{array}{c}
0.1603\\0.7543\\0.6366\end{array} \right],
\left| {{B_{4,2}}} \right\rangle  = \left[ \begin{array}{c}
 - 0.2879\\ - 0.5812\\0.7612\end{array} \right],\nonumber\\
&\left| {{B_{4,3}}} \right\rangle  = \left[ \begin{array}{c}
0.9441\\ - 0.3053\\0.1240\end{array} \right],
\left| {{B_{5,1}}} \right\rangle  = \left[ \begin{array}{c}
0.9735\\0.2148\\0.0786\end{array} \right],\nonumber\\
&\left| {{B_{5,2}}} \right\rangle  = \left[ \begin{array}{c}
 - 0.2030\\0.6536\\0.7291\end{array} \right],
\left| {{B_{5,3}}} \right\rangle  = \left[ \begin{array}{c}
 - 0.1052\\0.7257\\ - 0.6799\end{array} \right].\nonumber
\end{align*}
\end{subequations}

\end{document}